\documentclass[twocolumn,prb]{revtex4}
\usepackage{graphicx}
\usepackage{dcolumn}
\usepackage{bm}
\usepackage{amsmath}

\newcommand{\e}[1]{\mathrm{e}^{#1}}

\begin{document}

\preprint{}
\title{Anomalous magnetic transport in ferromagnetic graphene junctions}
\author{Takehito Yokoyama$^{1}$ and Jacob Linder$^{2}$}
\affiliation{$^1$Department of Physics, Tokyo Institute of Technology, 2-12-1 Ookayama, Meguro-ku, Tokyo 152-8551, Japan \\
$^{2}$Department of Physics, Norwegian University of Science and Technology, N-7491 Trondheim, Norway }
\date{\today}

\begin{abstract}
We investigate magnetotransport in a ferromagnetic/normal/ferromagnetic graphene junction where a gate electrode is attached to the normal segment. It is shown that the charge conductance can be maximal at an antiparallel configuration of the magnetizations. Moreover, we demonstrate that both the magnitude and the sign of the spin-transfer torque can be controlled by means of the gate voltage in the normal segment. In this way, the present system constitutes a \textit{spin-transfer torque transistor}.
These anomalous phenomena are attributed to the combined effect of the exchange field and the Dirac dispersion of graphene. 
Our prediction opens up the possibility of moving domain walls parallel or antiparallel to the current in a controllable fashion by means of a local gate voltage. 

\end{abstract}

\maketitle



%

%



Since its discovery, graphene has received much attention due to its rich potential from a fundamental and applied physics point of view. \cite{Ando,Katsnelson,Castro}
Graphene has a two-dimensional honeycomb network of carbon atoms and, as a result, electrons in graphene are governed by Dirac equation. 
The progress of practical
fabrication techniques for single graphene sheets has
allowed experimental study of this system, which has attracted a
tremendous interest from the scientific community.\cite{novoselov,zhang,novoselov_nature}

 Graphene is suitable for spintronics
applications: it exhibits gate-controlled
carrier conduction, high field-effect mobilities and a small spin-orbit interaction.\cite{Kane,Hernando}
Recent experiments on spin injection in
single layer graphene show a rather long spin flip length 
$ \approx 1 \mu$m at room temperature. \cite{Tombros}
 Spin injection into a graphene thin film has been successfully demonstrated by using non-local magnetoresistance measurements.\cite{Tombros,Ohishi,Cho} 
 The possibility of inducing ferromagnetic correlations in graphene due to the proximity effect by magnetic gates in close proximity to graphene has also been proposed.\cite{Haugen} In these works, ferromagnetism is induced by extrinsic means.

On the other hand, it is also possible to induce ferromagnetism intrinsically in graphene.\cite{Yazyev}  It has recently been predicted
that zigzag edge graphene nanoribbons become half-metallic under the influence of an external transverse electric field due to the different chemical potential shift at the edges,\cite{Son,Kan} which indicates the possibility of high controllability of ferromagnetism in graphene. In view of these, the study of spin transport in ferromagnetic graphene has recently developed. \cite{Haugen,Yokoyama,Yazyev2,Linder,Dell'Anna,Li}

One fundamental spintronic effect is the so-called spin-transfer torque.\cite{Tatara} When a spin-current is injected into a ferromagnetic layer with a magnetization misaligned compared to the polarization of the spin-current, 
the spin angular momentum of the injected electron changes upon entering and propagating through the ferromagnetic region since it follows the direction of the magnetization via $s-d$ interaction. This process gives a torque on the magnetization, and is called spin-transfer torque. 
Consequently, the magnetization can precess \cite{Slonczewski,Tsoi} or even be switched\cite{Slonczewski2,Berger,Myers,Katine}. 
Also, when a spin current is injected into a domain wall, spin-transfer torque can cause domain wall motion.\cite{Berger2} Finding methods to experimentally exert control over the spin-transfer torque following from spin-injection constitutes an important problem in the field of spintronics.

In this Rapid Communication, we study magnetotransport in a ferromagnetic/normal/ferromagnetic graphene junction where a gate electrode is attached to the normal segment. Firstly, we find that the charge-conductance can feature a maximum at an antiparallel configuration of the magnetizations, in contrast to the conventional spin-valve effect.
Moreover, we demonstrate how it is possible to \textit{control the magnitude and sign of the spin-transfer torque} by means of a local gate voltage, i.e. a transistor device for spin-transfer torque.
These results indicate that the combined effect of the exchange field and 
 the peculiar electric structure of graphene offers an unrivaled controllability of spin-transfer torque by means of a gate electrode.

Let us now explain the formulation. 
The fermions around Fermi level in graphene obey a massless relativistic Dirac equation at low energies. \cite{wallace}
\begin{figure}[htb]
\begin{center}
\scalebox{0.4}{
\includegraphics[width=20.0cm,clip]{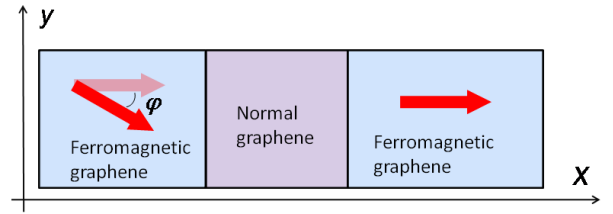}}
\end{center}
\caption{(color online) Schematic of the model: a graphene sheet where ferromagnetism is induced by a proximate host material. The left layer is assumed to have a soft magnetization, thus tunable in direction, whereas the magnetization of the right layer is fixed.} \label{f1}
\end{figure}
We consider a two dimensional ferromagnetic/normal/ferromagnetic graphene junction where a gate electrode is attached to the normal segment of the graphene sheet. See Fig. \ref{f1} for the schematic of the model. 
 The interfaces are parallel to the $y$-axis and located at $x=0$ and $x=L$. We consider the Hamiltonian 
 \begin{align}
 H =v_F \bm{k} \cdot \bm{\tau}  \otimes \sigma _0  + \tau _0  \otimes \bm{h} \cdot \bm{\sigma}  - E_F
 \end{align}
 in the ferromagnetic graphenes with ${\bm{h}} = h(\cos \varphi , - \sin \varphi ,0)$ for $x<0$ and ${\bm{h}} = (h, 0 ,0)$ for $x > L$ while in the normal graphene $H = v_F \bm{k} \cdot \bm{\tau}  \otimes \sigma _0  + V - E_F$. Here, $\tau$ and $\sigma$ are Pauli matrices in valley and spin space, respectively, $E_F$ is the Fermi energy, $V$ is the chemical potential shift in the normal region which is tunable by the gate voltage, and $h = \left| {\bm{h}} \right|$ is the magnitude of exchange field. 
One can use a magnet with very strong easy axis anisotropy for the right proximity ferromagnet and a soft magnet for the left ferromagnet which can be controlled by a weak external magnetic field.
Then, the wavefunctions are given as follows. In the left magnetic region, they read 
\begin{align}
\psi&(x\leq0) = \frac{1}{2}[1,\e{i\theta_\sigma},\sigma\e{-i\varphi},\sigma\e{i(\theta_\sigma-\varphi)}]\e{i k_\sigma\cos\theta_\sigma x}\notag\\
&+\frac{r_\sigma}{2}[1,-\e{-i\theta_\sigma},\sigma\e{-i\varphi},-\sigma\e{-i(\theta_\sigma+\varphi)}]\e{-ik_\sigma\cos\theta_\sigma x} \notag\\
&+\frac{r_{\bar{\sigma}}}{2}[1,-\e{-i\theta_{\bar{\sigma}}}, -\sigma\e{-i\varphi},\sigma\e{-i(\theta_{\bar{\sigma}}+\varphi)}]\e{-ik_{\bar{\sigma}}\cos\theta_{\bar{\sigma}}x}.
\end{align}
In the normal segment of graphene, we have:
\begin{align}
\psi&(0\leq x\leq L) = a[1,\e{i\theta'},0,0]\e{ik'\cos\theta'x}\notag\\
&+b[1,-\e{-i\theta'},0,0]\e{-ik'\cos\theta'x}+c[0,0,1,\e{i\theta'}]\e{ik'\cos\theta'x}\notag\\
&+d[0,0,1,-\e{-i\theta'}]\e{-ik'\cos\theta'x},
\end{align}
whereas finally in the right magnetic region:
\begin{align}
\psi(L\leq x) &= \frac{t_\sigma}{2}[1,\e{i\theta_\sigma},\sigma,\sigma\e{i\theta_\sigma}]\e{i k_\sigma\cos\theta_\sigma x} \notag\\
&+ \frac{t_{\bar{\sigma}}}{2}[1,\e{i\theta_{\bar{\sigma}}},-\sigma,-\sigma\e{i\theta_{\bar{\sigma}}}]\e{ik_{\bar{\sigma}}\cos\theta_{\bar{\sigma}} x}.
\end{align}
with angles of incidence $\theta_\sigma $ and $\theta'$, $\bar \sigma=-\sigma$, $\sigma=\pm$, $k_\sigma = (E + \sigma h + E_F )/v_F$ and $k'  = (E + E_F  + V)/v_F$. 
Because of the translational symmetry along the $y$-direction, the momentum parallel to the $y$-axis is conserved: 
$k_y= k_\sigma \sin \theta_\sigma  = k'\sin \theta '$.

By matching the wave functions at the interfaces $x = 0$
and $x = L$, we obtain the coefficients in the wavefunctions. The expressions of the coefficients are rather complicated and thus omitted here.  
The dimensionless charge conductance and spin-transfer torques, $G_c$ and $\tau_s^i$ $(i=x,y,z)$, are given by 
\begin{eqnarray}
G_c  = \frac{1}
{2}\sum\limits_\sigma  {\int_{ - \pi /2}^{\pi /2} {d\theta _\sigma  \psi ^\dag  j_c \psi } }, \\
\tau_s^i  = \frac{1}
{2}\sum\limits_\sigma  {\int_{ - \pi /2}^{\pi /2} {d\theta _\sigma  \psi ^\dag  j_s^i \psi } } 
\end{eqnarray}
with  $j_c  = \tau _x$ and $j_s^i  = \tau _x  \otimes \sigma _i $.

The total torque exerted on the right ferromagnetic layer should be obtained by the difference of the spin-currents deep inside the right and left layers. However, for good interfaces, to a very good approximation, no transverse angular momentum flows away from the interface, meaning that the torque is absorbed at or very near the interface. \cite{Ralph} For the present system, we have checked numerically that the vast majority of the torque is indeed absorbed at the interface region.

Below, we investigate charge conductance and spin-transfer torques at the Fermi level, i.e. $E=0$, and set $k_F L=1$ where $E_F=v_F k_F$. 
Spin-transfer torques depend on the position, and we here evaluate them at $x=L$. 
Their component parallel to the magnetization of the right ferromagnet cannot excite a precession of the magnetization, and hence we focus on their components parallel to the interface: $\tau_s^y$ and $\tau_s^z$. 

\begin{figure*}[htb]
\begin{center}
\scalebox{0.40}{
\includegraphics[width=45.0cm,clip]{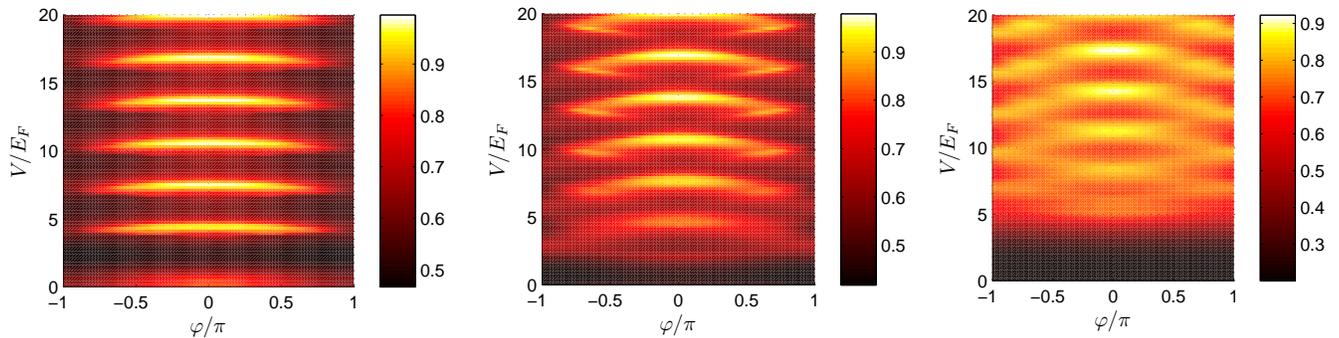}}
\end{center}
\caption{(color online) Normalized charge-conductance $G_c$ as a function of $\varphi$ and $V$ for $h/E_F=0.5$, 2, and 5 from left to right.} \label{f2}
\end{figure*}

\begin{figure*}[htb]
\begin{center}
\scalebox{0.4}{
\includegraphics[width=45.0cm,clip]{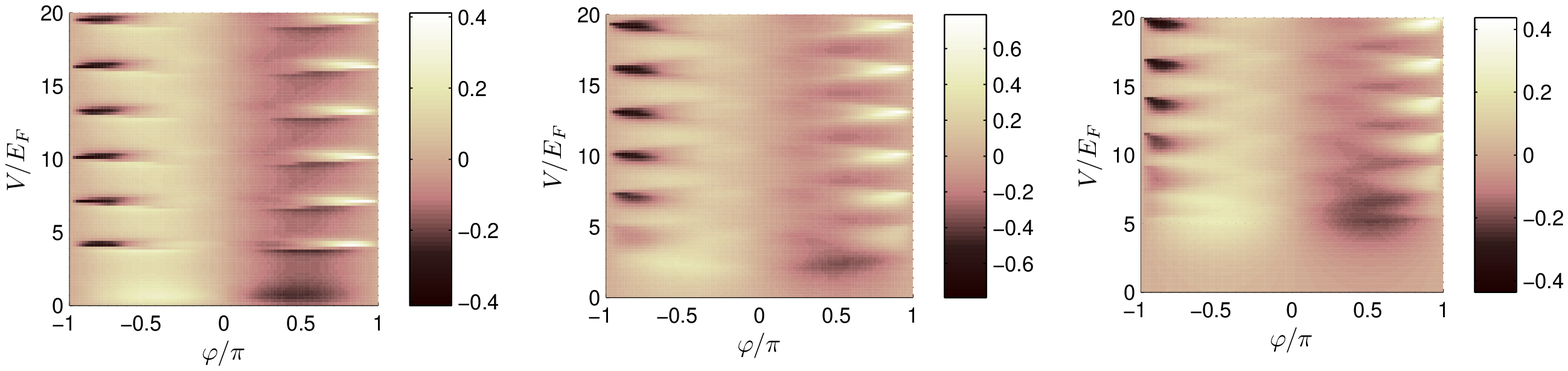}}
\end{center}
\caption{(color online) Normalized spin-transfer torque $\tau_s^y$ as a function of $\varphi$ and $V$ for $h/E_F=0.5$, 2, and 5  from left to right.} \label{f3}
\end{figure*}

\begin{figure*}[htb]
\begin{center}
\scalebox{0.4}{
\includegraphics[width=45.0cm,clip]{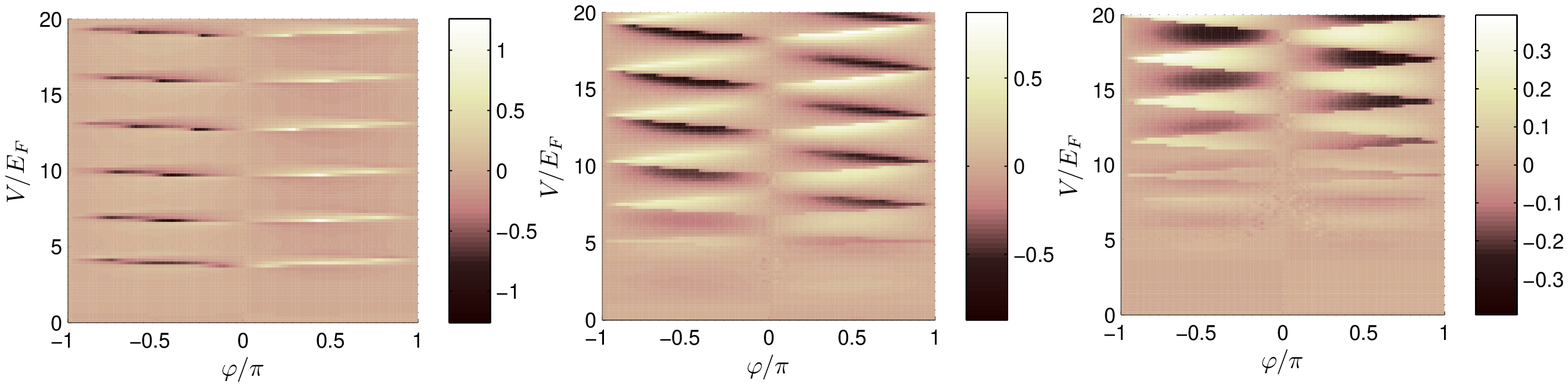}}
\end{center} 
\caption{(color online) Normalized spin-transfer torque $\tau_s^z$ as a function of $\varphi$ and $V$ for $h/E_F=0.5$, 2 and 5 from left to right.} \label{f4}
\end{figure*}

 Figure \ref{f2} shows the conductance $G_c$ as a function of $\varphi$ and $V$ for $h/E_F=0.5$, 2 and 5  from left to right. As seen, the conductance is an even function of $\varphi$. For small $h$, it becomes maximum at parallel configuration $\varphi=0$. 
However, interestingly, for large $h$, it attains a maximum near $\varphi=\pi$ at certain values of $V$. In graphene junctions, the transmission probability is not a decreasing function of the barrier height but oscillates with it (Klein tunneling).\cite{katsnelson} 
In fact, we find from the analytical expression of the spin-transfer torque that in the limit of $V \to \infty$, $L \to 0$, and $V L=Z=$ const., the barrier potential enters into the expression only in the form of $\sin Z$ and $\cos Z$. Thus, the spin-transfer torque oscillates with $Z$.
Therefore, the transmission probability at $\varphi=\pi$ can be larger than that at $\varphi=0$. 
This gives the anomalous magnetoconductance: the conductance at an antiparallel configuration can be larger than that at a parallel configuration. 
Similar anomalous magnetoconductance is also seen in ferromagnetic topological insulator junctions, although it is due to a different mechanism.\cite{Yokoyama2} The Klein tunneling in the present system also gives rise to oscillations in the conductance as a function of $V$\cite{katsnelson}.

 Figure \ref{f3} displays the corresponding plot of the spin-transfer torque $\tau_s^y$, which is seen to be an odd function of $\varphi$. 
For nonzero $\varphi$, the spin-transfer torque becomes nonzero and displays oscillations by varying $V$. Interestingly, we find that \textit{the sign of the torque can be changed by changing the gate voltage.}
 A similar plot of spin-transfer torque $\tau_s^z$ is shown in Fig. \ref{f4}, and is seen to behave qualitatively similar to $\tau_s^y$. 

The oscillation of the spin-transfer torque by the gate voltage, including its sign change, is unique to graphene because the origin of this behavior is the combined effect of the exchange field and Klein tunneling. Due to the exchange field, electrons for up and down spins effectively feel different barrier potential at the barrier region (the normal segment of graphene). Since the transmission probability oscillates with the barrier strength due to Klein tunneling, the 
transmission probability for spin-up and spin-down electrons oscillates with a shifted 
phase.\cite{Yokoyama} Thus, the spin current and hence the spin-transfer torque oscillates with the barrier strength or gate voltage.

A possible application of this result pertains to the fact that spin-transfer torque can be used to drive domain-wall motion in ferromagnetic materials\cite{Berger2,Tatara}, where the sign of the torque dictates the direction of propagation.
Since the sign of the spin-transfer torque in our system is controlled by a local gate electrode, this opens up the possibility of moving domain walls parallel or antiparallel to the current in a controllable fashion.

For the realization of our prediction of controllable spin-transfer torque by means of a gate voltage, $V/E_F > 1$ is required.
If we choose normal graphene with $k_F L=1$ and $E_F \simeq $ 1-10 meV, we need a junction length of around 0.1-1 $\mu$m. 
These values can be achieved by the present experimental techniques.
In order to realize ferromagnetic graphene, one may utilize the proximity effect by attaching magnetic insulator EuO to graphene. The estimated induced exchange field is of the order of a few meV. \cite{Haugen} State-of-the-art experimental techniques have recently made it possible to measure quantitatively the spin-transfer torque and its bias-dependence in conventional metallic tunneling junctions \cite{sankey_nphys_08, kubota_nphys_08}. If applied to the present graphene system, such methods could be utilized to observe the effects predicted in this paper. Finally, it should be noted that whereas graphene systems grown on substrates mostly belong to the diffusive regime of transport and also feature charge inhomogeneities, suspended graphene demonstrates ballistic properties for the charge-carriers and reduces inhomogeneities by orders of magnitude \cite{du}. Although the scattering formalism employed here assumes ballistic transport, we expect that the tunable spin-transfer torque remains qualitatively the same in the diffusive regime of transport.

In summary, we have studied magnetotransport in ferromagnetic/normal/ferromagnetic graphene junctions where a gate electrode is attached to the normal segment.
It is found that the charge-conductance can display a maximum at an antiparallel configuration of the magnetizations. 
Also, it is clarified that the spin-transfer torque oscillates with the gate voltage. In particular, a sign change of spin-transfer torque can be realized by means of the local gate voltage, meaning that a graphene spin-valve can act as a \textit{spin-transfer torque transistor} device.
These anomalous phenemena are attributed to the combined effect of the exchange field and Klein tunneling, a unique feature pertaining to graphene.
This opens up the possibility of moving domain wall parallel or antiparallel to the current in a controllable fashion by adjusting the gate voltage in the normal segment of the graphene spin-valve.


%


\end{document}